# Tailor the functionalities of metasurfaces: From perfect absorption to phase modulation

Che Qu[1,$], Shaojie Ma[1,$], Jiaming Hao[2,*], Meng Qiu[1], Xin Li[1], Shiyi Xiao[1], Ziqi Miao[1], Ning Dai[2], Qiong He[1,3], Shulin Sun[4], Lei Zhou[1,3,*]

[1] *State Key Laboratory of Surface Physics, Key Laboratory of Micro and Nano Photonic Structures (Ministry of Education) and Physics Department, Fudan University, Shanghai 200433, China*

[2] *National Laboratory for Infrared Physics, Shanghai Institute of Technical Physics, Chinese Academy of Science, Shanghai 200083, China*

[3] *Collaborative Innovation Center of Advanced Microstructures, Fudan University, Shanghai 200433, China*

[4] *Shanghai Engineering Research Center of Ultra-Precision Optical Manufacturing, Green Photonics and Department of Optical Science and Engineering, Fudan University, Shanghai 200433, China*

[$] These authors equally contributed to this work

[*] Corresponding authors: Lei Zhou, phzhou@fudan.edu.cn; Jiaming Hao, jiaming.hao@mail.sitp.ac.cn.

**Abstract:**

Metasurfaces in metal/insulator/metal configuration have recently been widely used in photonics research, with applications ranging from perfect absorption to phase modulation, but why and when such structures can realize what kind of functionalities are not yet fully understood. Here, based on a coupled-mode theory analysis, we establish a complete phase diagram in which the optical properties of such systems are fully controlled by two simple parameters (i.e., the intrinsic and radiation losses), which are in turn dictated by the geometrical/material parameters of the underlying structures. Such a phase diagram can greatly facilitate the design of appropriate metasurfaces with tailored functionalities (e.g., perfect absorption, phase modulator, electric/magnetic reflector, etc.), demonstrated by our experiments and simulations in the Terahertz regime. In particular, our experiments show that, through appropriate structural/material tuning, the device can be switched across the functionality phase boundaries yielding dramatic changes in optical responses. Our discoveries lay a solid basis for realizing functional and tunable photonic devices with such structures.

## 1. Introduction

Metasurfaces are man-made ultrathin photonic materials (much thinner than working wavelength) constructed by subwavelength planar units with desired electromagnetic (EM) properties, which exhibit unusual abilities to manipulate EM waves [1]. Recently, metasurfaces in metal/insulator/metal (MIM) configuration have been widely used in photonic research [2-18]. Such structures typically consist of a layer of planar metallic resonators and a continuous metallic film separated by a dielectric spacer (see Fig. 1(a)). Based on MIM metasurfaces, a variety of wave-manipulation effects have been realized in a wide frequency range, including perfect absorption [2-5], polarization control [6-8], phase modulation [9-11], anomalous light reflection [12-14], flat-lens focusing [15, 16], and holograms [17, 18]. It is intriguing to see that MIM metasurfaces can behave distinctly under slight structural tuning and thus can realize such diversified applications. However, whereas full-wave simulations can well reproduce the discovered phenomena on such systems [2-20], the inherent physics underlying these distinct effects remains obscure. In particular, why and with what parameters can these MIM metasurfaces exhibit certain functionalities (e.g., perfect absorption, phase modulation, etc.) are still not fully understood, despite of some initial attempts [21, 22].

The key motivation of this paper is to answer these questions and to provide a universal guidance for designing MIM metasurfaces with tailored functionalities. To achieve this end, we first employ the coupled-mode theory (CMT) [23-25] to derive a generic phase diagram in which the functionality of a MIM metasurface is governed by two simple lumped parameters (Sec. 2). These two parameters, describing respectively the radiation loss and intrinsic loss of the MIM structure, are linked with the system's structural/material details through two relationships that can be derived analytically under certain conditions (Sec. 3). The derived formulas guide us to experimentally realize a series of Terahertz (THz) metasurfaces with distinct (yet well controlled) functionalities (Sec. 4). In particular, functional phase transitions were experimentally demonstrated via tuning the structural/material parameters of the metasurfaces, again assisted by the phase diagram and the analytical relationships (Secs. 4-5). Our results not only set up a unified platform to understand all previous studies related to MIM metasurfaces [2-22], but more importantly, also provides a powerful tool to guide the future design and realization of functional complex meta-devices with unusual properties (e.g., active tunability, etc.).

## 2. A generic phase diagram for MIM metasurfaces

The simplest model to describe the complex MIM structures (Fig. 1(a)) is the single-port resonator as schematically depicted in Fig. 1(b). Near-field coupling between two metallic layers can form a resonance in the structure at a frequency $\omega_0$ [26-28]. Meanwhile, presence of a metallic ground plane in such a structure ensures that no EM wave can pass through it, so that only one port (i.e., the reflection port) of the resonator needs to be considered (Fig. 1(b)). According to CMT [23-25], the (complex) reflection coefficient $r$ of a one-port single-resonance model can be derived as

$$r = -1 + \frac{2/\tau_r}{-i(\omega - \omega_0) + 1/\tau_a + 1/\tau_r} \tag{1}$$

where $\tau_a$ and $\tau_r$ are two parameters denoting the life times of the excited resonance due to absorption inside the structure and radiation to the far field, respectively. In what follows, we adopt two dimensionless parameters, $Q_a = \omega_0 \tau_a / 2$ and $Q_r = \omega_0 \tau_r / 2$, describing respectively the absorptive and radiative quality factors, to make our discussions more generic.

Equation (1) shows that the physical property of our resonator is fully determined by $Q_a$ and $Q_r$. A simple calculation reveals that all previously reported behaviors of MIM structures [2-22] can be easily produced by tuning these two parameters. Figure 1(c) depicts how the absorbance of our model at its resonance frequency $\omega_0$, calculated by $A = 1 - |r|^2$, varies against $Q_a$ and $Q_r$. We find that absorbance $A$ depends sensitively on $Q_a$ and $Q_r$. While perfect absorption $A=1$ happens when $Q_a = Q_r$ (see also [29, 30]), the absorbance decreases significantly as leaving this phase boundary (Fig. 1(c)). Figure 1(e) compares the calculated spectra of $|r|^2$ for 5 typical systems, selected from an equi-$Q$ line (see Fig. 1(c)) across the phase boundary defined by $Q_a = Q_r$, where the total $Q$ factor is defined by $Q = Q_a Q_r / (Q_a + Q_r)$. However, while the $|r|$ spectrum for $Q_a = Q_r$ does show a pronounced dip at $\omega = \omega_0$ in consistency with the perfect absorption, the difference between two regions separated by this phase boundary *cannot* be clearly seen from the $|r|^2$ spectra alone.

The two phase regions (i.e., $Q_a < Q_r$ and $Q_a > Q_r$) can be unambiguously separated as we check the reflection phase ($\phi$) spectra of the 5 systems depicted in Fig. 1(f). In the case of $Q_a > Q_r$, the reflection phase $\phi$ always undergoes a continuous $-180°$ to $180°$ variation as frequency passes through the resonance and the fact that $\phi = 0°$ at resonance indicates that such a resonance is a "magnetic" one [26-28]. On the contrary, In the cases of $Q_a < Q_r$, the variation of $\phi$ only occupies a small range less than $180°$. In particular, the fact that $\phi = 180°$ at $\omega = \omega_0$ reveals the "electric" nature of the involved resonance. These results suggest that the phase variation range $\Delta\phi$ of the $\phi$ spectrum is another indicator to label the physical behaviors of the system, which is more clearly shown in Fig. 1(d). Explicitly, in the $Q_a > Q_r$ region where the intrinsic absorption is weak, the resonator is under-damped and $\Delta\phi$ can approaches the full $360°$ range, so that the device can realize phase-modulation-related effects such as polarization control and anomalous reflection [6-8, 13]. On the other hand, when the resonator is located in the $Q_a < Q_r$ region, it becomes over-damped and can only provide very limited phase modulation freedom.

The rich physics behind the phase diagrams shown in Fig. 1(c)-(d) can be better understood based on the Smith curves --- the evolution of $r$ in complex plane as frequency varies from 0 to $\infty$. As shown in the inset to Fig. 1(f), all curves start and end at similar points close to $r = -1$ (see also Eq. (1)), but at resonance, they cross the real axis at different points determined by $r = (Q_a - Q_r)/(Q_a + Q_r)$. Therefore, the ratio between $Q_a$ and $Q_r$ directly determines whether the Smith curve enclose the origin or not. In the cases of $Q_a < Q_r$, the corresponding Smith curves do not enclose the origin so that the span of phase variation $\Delta \phi$ is less than $180°$. However, in the cases of $Q_a > Q_r$, $\Delta \phi$ can approach to $360°$ since the Smith curves enclose the origin in such cases. Finally, the smith curve passes through the origin the case of $Q_a = Q_r$, which is corresponding to the perfect absorption situation [29, 30]. The competitions between two $Q$ factors thus generate a variety of physical effects, which have been separately discovered previously based on MIM systems [2-18].

## 3. Analytical relationships between *Q* factors and real geometrical/material parameters

Having understood the crucial roles played by $Q_a$ and $Q_r$ in determining the functionality of a MIM metasurface, we now study the relationships between these two parameters and the configuration details of realistic systems. Such relationships can serve as a bridge to help design MIM metasurfaces with tailored functionalities. Here we derive two analytical formulas to establish such relationships for an MIM structure with the simplest geometry. These formulas can not only immediately guide us to realize functional MIM metasurfaces with simplest geometry (Sec. 4), but more importantly, also shed light on the design of general metasurfaces with more complex geometries (Sec. 5).

The schematics of our model system is shown in the inset to Fig. 2(a), where the top metallic layer consists of air slits (with widths *a*) arranged with a lattice constant *d*. The thicknesses of the metallic and dielectric layers are $h_m$ and $h$, respectively. The scattering properties of such a model system, under the illumination of a normally incident plane wave with $\vec{E} \parallel \hat{x}$, can be rigorously studied using the mode-expansion theory (MET) [31-33]. For structures under subwavelength ($d \ll \lambda, h < \lambda$) and thin-slit ($a \ll d$) conditions, a simple analytical formula can be derived (see Supporting Information) for the radiation *Q* factor of such a system

$$Q_r = \frac{1}{2k_0 h} \cdot \sum_m \Gamma(m) \cdot \frac{\sin^2(m\pi a/d)}{(m\pi a/d)^2}, \qquad (2)$$

where $\Gamma(m) = \varepsilon k_0^2 \cdot ((2m\pi/d)^2 + \varepsilon k_0^2)/((2m\pi/d)^2 - \varepsilon k_0^2)^2$ is a dimensionless parameter to weight the contribution of the *m*-th internal mode inside the cavity, $\varepsilon$ denotes the relative permittivity of the dielectric spacer, and $\sin^2(m\pi a/d)/(m\pi a/d)^2$ describes the coupling strength between the external field and the *m*-th internal mode. Formula (2) is obtained by rigorously calculating $Q_r$ as the ratio of the energy stored inside the resonator to the energy radiated outside the resonator per time-oscillation cycle, based on the field distribution calculated by the MET (see Supporting Information). Note that we have purposely assumed that both metals and dielectrics are lossless when deriving Eq. (2), since absorptions are explicitly considered when calculating another parameter $Q_a$. Interestingly, $Q_r$ is found to be inversely proportional to the cavity thickness *h*, which is a *universal* behavior of such MIM structure. Such an intriguing effect plays a crucial role in our further studies, and it can be attributed to the near-field coupling between two metallic layers. As $h \to 0$, near-field coupling is significantly enhanced resulting in stronger local fields inside the cavity, and in turn, an enlarged $Q_r$ factor of the resonator.

We now consider $Q_a$, which is defined as the ratio between the total energy

$U$ stored inside the cavity and the energy absorbed by the dielectric spacer $P_d$ and by the metals $P_m$, during a time-oscillation circle. While an analytical expression of $Q_a$ can also be derived based on the MET (see Supporting Information), here we present a simple and intuitive derivation assuming that the field distributes uniformly inside the cavity. Obviously, $U$ is proportional to the total volume of the cavity and thus $U \propto \text{Re}(\varepsilon) \times h \cdot d$. Meanwhile, $P_d$ should be proportional to $\text{Im}(\varepsilon)\omega_0 \times h \cdot d$ since the absorption happens everywhere inside the dielectric medium where $\text{Im}(\varepsilon)$ is non-zero. Finally, considering the exponential decay of electric field inside a metal layer $E \sim e^{-z/\delta}$ with $\delta = \sqrt{2/\mu_0\omega_0\sigma_m}$ being the skip-depth of the metal, we find that metallic absorption should be proportional to $\text{H}(\delta) = \delta\left(1 - e^{-2h_m/\delta}\right)$, which is the effective field-decaying length. Collecting all these considerations, we finally have

$$Q_a = \frac{\omega_0 U}{P_d + P_m} = \frac{\text{Re}(\varepsilon)hd}{\alpha \text{Im}(\varepsilon)hd + \beta(2d-a)\text{H}(\delta)} \quad , \quad (3)$$

where $\alpha$ and $\beta$ are two parameters to define the relative contributions from the dielectric and metallic media, respectively. Interestingly, we note that Eq. (3) can actually be derived from the rigorous MET, from which the detailed expressions of $\alpha$ and $\beta$ are explicitly given (see Supporting Information).

**4. Tailoring metasurfaces through structural/material tuning**

Previous sections have revealed that, tuning the structural/materials details of a MIM metasurface can significantly modify its $Q_r$ and $Q_a$, and in turn, change the functionality of the system. Here we combine THz experiments and simulations to illustrate how the idea works.

The most important parameter to "tune" the property of a metasurface is the cavity thickness $h$. As shown in Eqs. (2-3), in the limit of $h \to 0$, while $Q_r$ scales inversely with respect to $h$, $Q_a$, on the other hand, almost linearly depends on $h$. Such distinct $h$ dependences of two $Q$ factors are more clearly shown in Fig. 2(a), where two lines representing $Q_r \sim h$ and $Q_a \sim h$ relations calculated by Eqs. (2-3), intersect at a critical point at $h_c \approx 7.5\,\mu m$. Combining Fig. 2(a) and Fig. 1, we immediately understand that decreasing $h$ can drive the system from an under-damped resonance ($Q_a > Q_r$) to an over-damped one ($Q_r > Q_a$), which is more explicitly illustrated in Fig. 2(b). Noting that the two regions have completely different EM functionalities (see Fig. 1(c)-(d)), we thus understand that tuning the cavity thickness can dramatically change the property of a metasurface, especially in the vicinity of the critical thickness $h \sim h_c$.

These predictions have been verified by our THz experiments and simulations. We fabricated a series of metasurfaces with different $h$ following the standard optical lithography procedures. An optical image of a typical sample is shown in the inset to Fig. 2(b). We next used a THz time-domain-spectroscopy (TDS) to measure the spectra of reflection amplitude/phase of these samples and plot the results in Fig. 2(c) and (d). The measured spectra are in good agreement with finite-difference-time-domain (FDTD) simulations on realistic structures. The distinct behaviors in phase spectra of different samples (Fig. 2(d)) already imply a phase transition from under-damped to over-damped resonance, as $h$ decreases from $11\,\mu m$ to $3\,\mu m$. As a quantitative check, we retrieved the values of two $Q$ factors by fitting the measured and simulated spectra of different samples to Eq. (1) (see Supporting Information), and plotted the obtained results in Fig. 2(a)-(b). Such retrieved values are in reasonable agreement with the model analytical results. In particular, the crossover between two functionality regions is clearly seen in Fig. 2(b).

The slit width $a$ is another critical parameter that can dramatically modify the physical properties of a MIM structure. The physics can again be easily understood based on two analytical formulas Eqs. (2-3). First, we note from Eq. (2) that $Q_r$ only weakly depends on $a$, which is verified by the computed $Q_r \sim a$ curve depicted in Fig. 3(a), for a model with $d = 100\,\mu\text{m}$ and $h = 8\,\mu\text{m}$ fixed. The physics is that near-field coupling (dictated by $h$) between two metallic layers plays much more important role than the lateral structural parameter in determining the radiative $Q$ factors of such double-layer systems. Meanwhile, decreasing $a$ can dramatically increase the contribution of metallic absorption through increasing the metal occupations in the system, and in turn, decrease $Q_a$ appreciably (see Eq. (3)). This is verified by the calculated $Q_a \sim a$ curve for the same model (Fig. 3(a)). The distinct $a$-dependences of two $Q$ factors creates a crossover point at $a \approx 14\,\mu\text{m}$. Therefore, a phase transition similar to the case discussed in Fig. 2(d) can be driven by varying the parameter $a$, as shown more clearly in Fig. 3(d). These predictions were again verified by our THz experiments and FDTD simulations for a series of MIM metasurfaces with different parameter $a$ (see Fig. 3(c)-(d)). Again, the retrieved $Q$ factors from experiment and FDTD results confirmed the physical picture established in last sections. We note that the discrepancy between analytical and FDTD calculations becomes large as $a$ increases (Fig. 3(a)-(b)), because Eqs. (2-3) are derived based on the thin-slit approximation ($a \ll d$). However, the overall trend and the phase-region crossover are not affected by such quantitative discrepancies.

Finally, we show that the functionality switching can also be realized by tuning the material properties, particularly the loss parameters of the constitutional media. It can be easily expected that increasing the conductivity $\sigma_\text{d}$ of the dielectric spacer and/or decreasing the conductivity $\sigma_\text{m}$ of the metal can significantly decrease $Q_a$ through enhancing the absorption, but they have relatively minor effects on $Q_r$ which is predominantly determined by the resonator's geometry. Therefore, tuning $\sigma_\text{d}$ and/or $\sigma_\text{m}$ can have dramatic effects on the device's performances. To illustrate how the idea works, we employed FDTD simulations to study the optical properties of a series of metasurfaces with varying $\sigma_d$ and then retrieved the two $Q$ factors from the calculated spectra. Clearly, increasing $\sigma_d$ indeed drives the resonator to move downward in the $Q_r - Q_a$ phase diagram (Fig. 4(a)), and a critical transition happens as $\sigma_d$ equals a particular value. Meanwhile, a similar loss-driven functionality switching is shown in Fig 4(b) for the case with varying $\sigma_m$, where the resonator moves upward in the $Q_r - Q_a$ phase diagram as $\sigma_m$ increases, because a larger $\sigma_m$ means a better metal and thus a larger $Q_a$.

These numerical results point out the possibility of making actively tunable devices based on the MIM metasurfaces. For example, one can use semiconductor as the dielectric space to construct a MIM device, and can tune the loss of the semiconductor via optical pumping [34] to drive the MIM resonator from an under-damped one to an over-damped one. Very significant active phase modulation can be achieved accompanying such a switching. Alternatively, we note that the conductivity of a graphene can be significantly tuned via electric gate control, which can also be applied to drive the above-mentioned functionality switching [35-38].

## 5. Extension to general situations with complex lateral geometry

Although the results presented in Secs.3-4 are based on a simple geometry, the conclusions drawn are general enough to work for more complex situations. In this section, we take metasurface with cross-like lateral pattern as an example to illustrate how the idea works. The geometry of this kind of metasurface is shown in Fig. 5(a), with $d$, $l$, and $w$ denoting the array period, bar length and bar width, respectively. The role of these parameters can be understood by mapping present structure to the stripe case studied in last sections. Obviously, the cavity thickness $h$ plays the same role as in previous cases, and the bar length $l$ is just corresponding to $(d-a)$ in the stripe pattern. However, bar-width $w$ does not have a clear counterpart in the stripe pattern, and thus has to be studied separately. It is well known that a planar resonator with thinner wires exhibits a higher $Q_r$, and thus decreasing $w$ must enhance the $Q_r$ of the system. Therefore, with the roles played by all three parameters known, we can utilize the conclusions drawn in Secs.3-4 to guide us design appropriate complex MIM structures with desired functionalities. Specifically, to drive an MIM structure to transit from the over-damped region to the under-damped region, one needs to reduce the $Q_r$ factor through increasing $h$ and/or $w$, and enhance the $Q_a$ factor through increasing $h$ and/or decreasing $l$.

This prediction has been verified by our THz experiments and FDTD simulations. We purposely designed a series of MIM structures with enhancing $h$, $w$ and diminishing $l$, and Fig. 5(a) shows the optical pictures of three typical samples that we fabricated. The measured spectra of reflection amplitude/phase (Fig. 5(c)-(d)) as well as their corresponding retrieved $Q$ factors (Fig. 5(b)) verified our conjecture. In principle, FDTD simulations demonstrate that "tuning" the parameters ($h$, $l$ and $w$) of a MIM structure can *continuously* change the optical properties of the structure, and drive the system from one functionality region to another (see Supporting Information). The three experimentally tested samples represent three typical cases, which are a magnetic reflector with under-damped resonance (sample S3), a perfect absorber (sample S2) and an electric reflector with over-damped resonance (sample S1).

## 6. Conclusions

In summary, we combine theory and experiments to establish a unified phase diagram to understand the diversified functionalities discovered on MIM metasurfaces, and provide practical approaches to "tune" the functionalities of such systems via structural/material changing. These generic results, independent of the frequency of interest, can guide people to design their own MIM metasurfaces with tailored and even tunable optical responses in different frequency domains, which may find numerous applications in photonic research.

## 7. Experimental Section

**Sample Fabrications**

Our samples were fabricated with the following procedures. We first deposited a 50 nm – thick Au layer with a 10 nm – thick adhesive Ti layer on a bare Si substrate using electron-beam evaporation and then spin-coated a polyimide layer with required thickness (multiple spin-coating is used when thicker space is required), and finally fabricated the designed metallic patterns (slits or crosses) on top of the polyimide layer by standard photolithographic method, followed by a lift-off process. All fabricated samples are 1.5 cm * 1.5 cm in size.

**THz Time-Domain Spectroscopy for Experimental Characterization**

We used a commercial TDS system (Zomega-3, Zomega Terahertz Corporation) to perform reflection measurements. The THz waves were generated from a GaAs emitter illuminated by a femtosecond fiber laser pulse train with a wavelength of 780 nm, a repetition rate of 80 MHz, and a pulse width of 125 fs. The horizontal electric field of the reflected THz wave was detected using a typical electro-optic sampling method (with a ZnTe crystal) in the time domain. The THz section of the setup was purged with dry $N_2$ gas to prevent the fingerprint absorption of THz waves in air. The spectral amplitude and phase were determined from the measured data by using numerical Fourier transformation.

**FDTD simulations**

FDTD simulations were performed using a numerical solver (CONCERTO 7.0, Vector Fields Limited, UK (2008)). In our simulations, we assumed that the relative permittivity of the dielectric medium (polyimide) is $\varepsilon_d = 3.1 + i\sigma_d/\omega$ with $\sigma_d = 8$ S/m, and treated Au as a lossy metal with conductivity $\sigma_m = 1.0 \times 10^6$ S/m.

**Supporting Information**

Supporting Information is available from the Wiley Online Library or from the author.

**Acknowledgements**

This work was supported by National Natural Science Foundation of China (Grant Nos. 11474057, 61471345, 11174055, 11204040, and 11404063), MOE of China (Grant No. B06011), and Shanghai Science and Technology Committee (Grant Nos. 14PJ1409500 and 14PJ1401200).

**Figure captions**

**Figure 1**

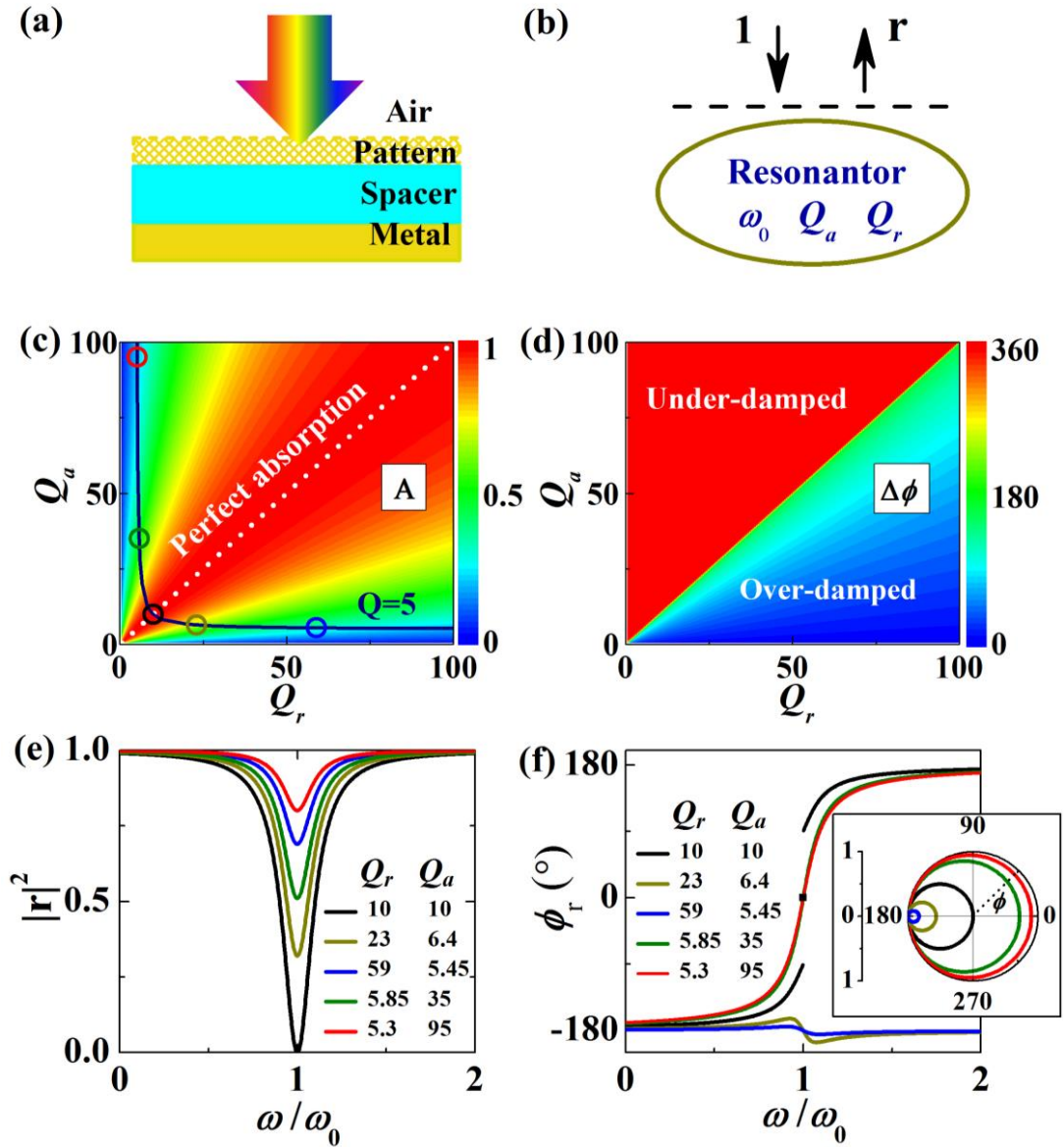

Fig. 1. Schematics of (a) the MIM system and (b) the single-port resonator model in CMT. Phase diagrams of (c) the on-resonance absorption A and (d) the span of reflection phase $\Delta\phi$ versus $Q_r$ and $Q_a$, calculated with CMT for the single-port model shown in (b). CMT-calculated spectra of (e) reflectance and (f) reflection phase for five systems represented by the open circles on the $Q=5$ contour line in (c). Inset to (f) depicts the Smith curves of the reflection coefficients of the five systems.

**Figure 2**

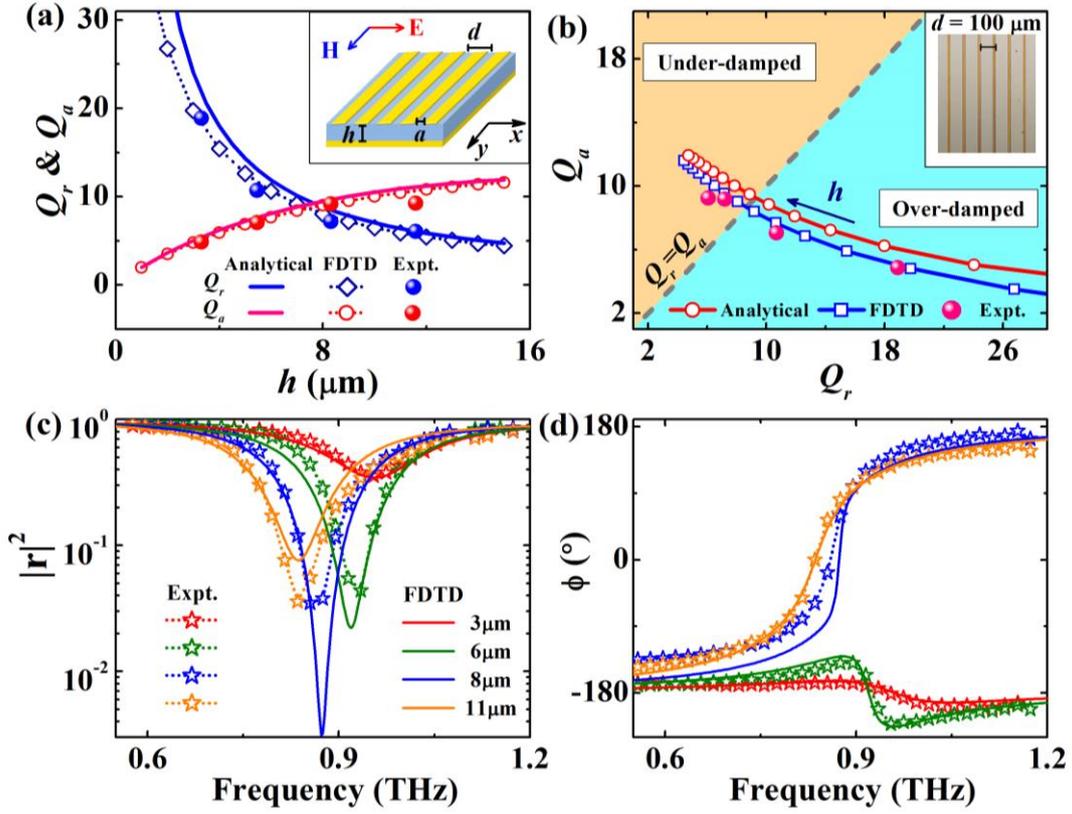

Fig. 2. (a) $Q_r$ and $Q_a$ for a series of metasurfaces (see inset for system geometry) with varying $h$, obtained by analytical calculations (lines) based on Eqs. (2-3), or retrieved from FDTD simulated spectra (open symbols) and from experimental results (solid symbols). Other geometrical parameters of these metasurfaces are fixed as $a = 20\,\mu\text{m}$, $d = 100\,\mu\text{m}$, $h_m = 0.05\,\mu\text{m}$. (b) Re-plot of the results displayed in (a) into a $Q_a - Q_r$ phase diagram, where the $Q_r = Q_a$ line separates the over-damped region (blue) and the under-damped region (orange). Inset shows an optical image of the sample with $h = 8\,\mu\text{m}$. Spectra of (c) reflectance and (d) reflection phase of four typical samples with $h$ given in the legend of (c), obtained by THz-TDS measurements (symbols) and FDTD simulations (lines).

**Figure 3**

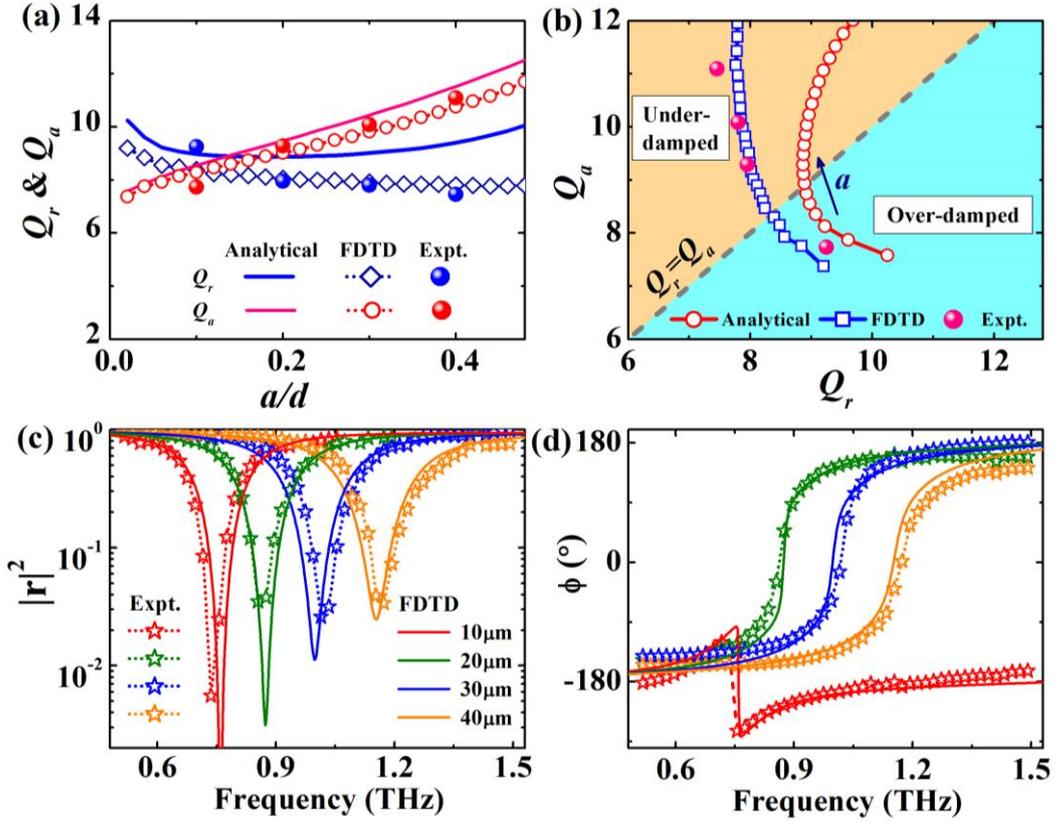

Fig. 3. (a) $Q_r$ and $Q_a$ for a series of metasurfaces with varying slit-width $a$, obtained by analytical calculations (lines) based on Eqs. (2-3), or retrieved from FDTD simulated spectra (open symbols) and experimental results (solid symbols). Other geometrical parameters of these metasurfaces are fixed as $h = 8\,\mu m$, $d = 100\,\mu m$, $h_m = 0.05\,\mu m$. (b) Re-plot of the results displayed in (a) in a $Q_a - Q_r$ diagram. Spectra of (c) reflectance and (d) reflection phase of four typical samples with $a$ given in the legend of (c), obtained by THz-TDS measurements (symbols) and FDTD simulations (lines).

**Figure 4**

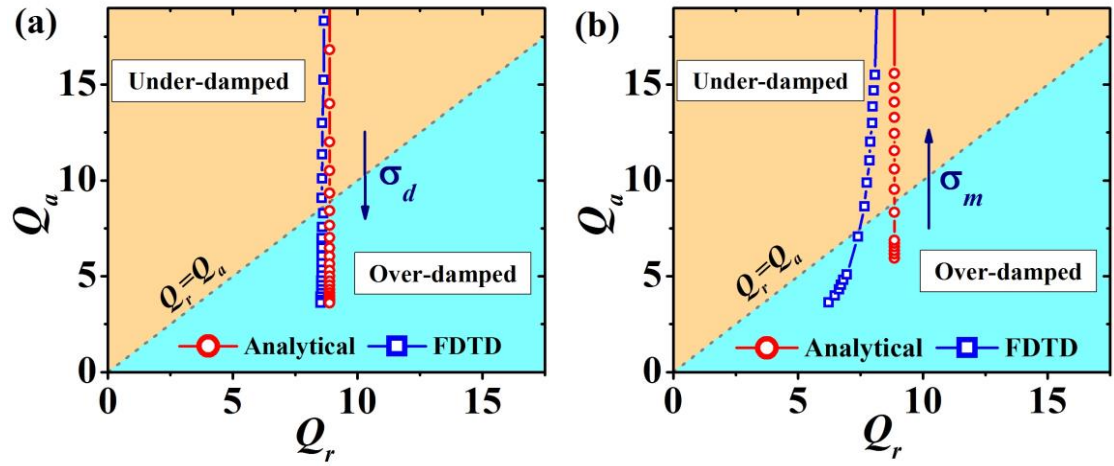

Fig. 4. (a) $Q_r$ and $Q_a$ for a series of metasurfaces with $\sigma_d$ varying from $8\,\text{S/m}$ to $50\,\text{S/m}$ and with metal assumed lossless, obtained by analytical calculations (circles) based on Eqs. (2-3), or retrieved from FDTD simulated spectra (squares). (b) $Q_r$ and $Q_a$ for a series of metasurfaces with $\sigma_m$ varying from $5\times 10^4\,\text{S/m}$ to $2\times 10^6\,\text{S/m}$ and with $\sigma_d$ fixed as $0\,\text{S/m}$, obtained by analytical calculations (circles) based on Eqs. (2-3), or retrieved from FDTD simulated spectra (squares). Other parameters are fixed as $h=8\,\mu\text{m}$, $a=20\,\mu\text{m}$, $d=100\,\mu\text{m}$, $h_m=0.05\,\mu\text{m}$.

**Figure 5**

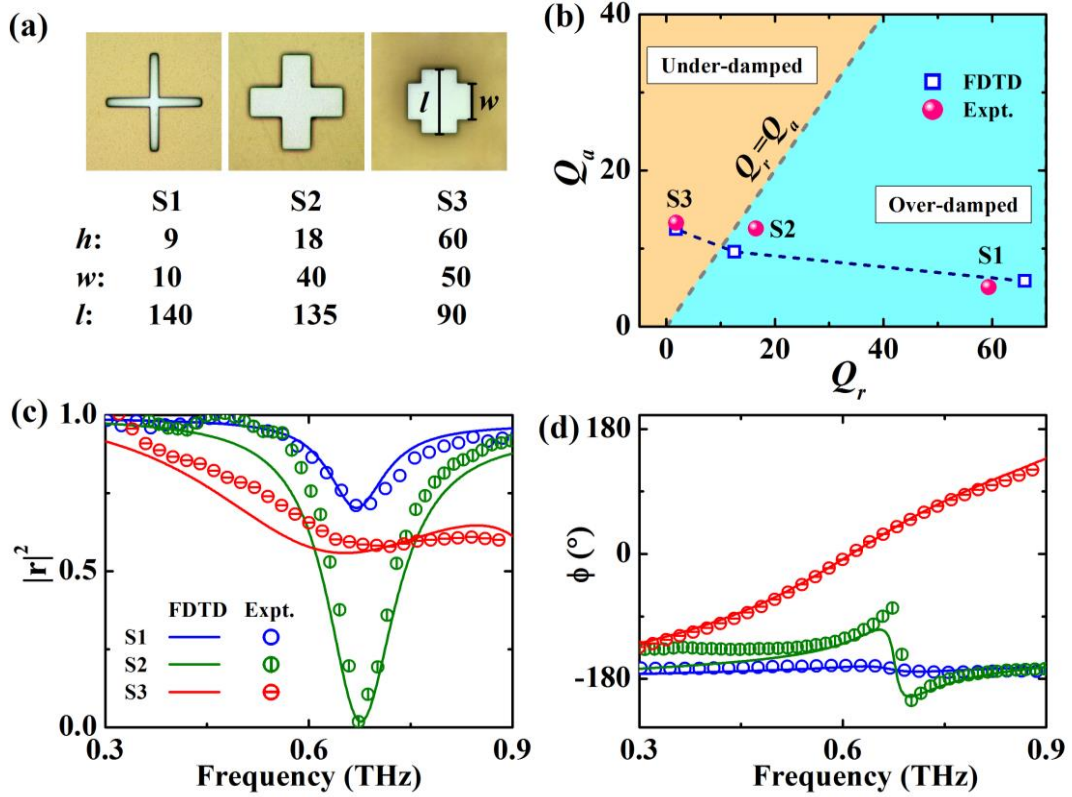

Fig. 5. (a) Optical images of three fabricated metallic cross-shaped samples (labeled by S1, S2, and S3) with geometrical parameters given in unit of μm. (b) $Q_r$ and $Q_a$ for three metasurfaces retrieved from FDTD simulated spectra (open squares) and retrieved from experimental results (solid symbols). Spectra of (c) reflectance and (d) reflection phase of three samples obtained by THz-TDS measurements (symbols) and FDTD simulations (lines).